\documentclass{article}
\usepackage{amssymb, amsmath, textcomp}
\usepackage{graphicx}
\usepackage{verbatim}   
\usepackage{color}      
\usepackage{subfigure}  
\usepackage{amsmath}    
\usepackage{epsfig}     
\usepackage{dsfont}
\usepackage{amsthm}
\usepackage{mathrsfs, caption}
\numberwithin{equation}{section}
\newtheorem*{thm4-2}{Theorem 4.2}
\newtheorem*{thm2-1}{Theorem 2.1}
\newtheorem*{thm3-1}{Theorem 3.1}
\newtheorem*{thm4-1}{Theorem 4.1}
\newtheorem*{thm5-1}{Theorem 5.1}
\newtheorem*{prop3-1}{Proposition 3.1}
\newtheorem*{prop3-2}{Proposition 3.2}
\newtheorem*{prop3-3}{Proposition 3.3}

\usepackage{mathtools, verbatim}

\makeatletter

\newcommand{\Rmnum}[1]{\expandafter\@slowromancap\romannumeral #1@}
\makeatother

\title{Parametric Solution Of Certain Nonlinear Differential Equations In Cosmology II}

\author{Jennie D'Ambroise\\
\small\it Bard College, Mathematics Program\\
\small\it P.O. Box 5000\\
\small\it Annandale-on-Hudson, NY 12504, USA\\
 \small \it  jdambroi@bard.edu\\
 \\
Floyd L. Williams\\
\small\it Department of Mathematics and Statistics\\ 
\small\it University of Massachusetts at Amherst\\
\small\it Amherst, MA 01003, USA\\
\small\it williams@math.umass.edu}
\date{}

\begin{document}
\maketitle

\begin{center}To the memory of A. E. Nussbaum\end{center}

\begin{abstract} 
This paper continues earlier work where an explicit parametrized solution of a particular nonlinear ordinary differential equation was obtained in terms of the Weierstrass elliptic phi-function, sigma, function, and zeta function -- work which therefore generalized that of G.Lemaitre.  Further study of this solution and applications to cosmology are presented.
\end{abstract}

\vspace{.25in}



\noindent {2010 Physics and Astronomy Classification Scheme (PACS): 02.30.Gp, 02.30.Hq,\\
02.30.Jr, 04.20.Jb, 04.40.Nr}

\section{Introduction}

\indent \indent  The Belgium Priest G. Lemaitre, in 1933, was the first to study spherically symmetric distributions of matter without pressure in a non-static, inhomogeneous cosmological model \cite{1}.  For such a model, being dust sourced, one has the Lemaitre-Tolman-Bondi metric (for spherical coordinates $(r, \theta, \phi)$)
\begin{equation}ds^2=dt^2-\frac{ \left( \frac{\partial R}{\partial r} \right)^2 }{ 1+2E(r)} dr^2 - R(t,r)^2\left(  d\theta^2+\sin^2\theta d\phi^2 \right).\end{equation}
Here $E(r)$ is an arbitrary function (a local curvature function) and $R(t,r)$ is both temporal and spatially dependent.  For $\dot{R}=\frac{\partial R}{\partial t}$, the  dynamical equation (or a first integral of the Einstein equations)
\begin{equation}\dot{R}^2=2E(r)+\frac{2M(r)}{R}+\frac{\Lambda}{3} R^2 \end{equation}
holds, where $M(r)$ is an arbitrary (mass) function and where $\Lambda $ is a cosmological constant.  Equation (1.2), which exhibits an energy interpretation of $E(r)$ as well, was solved parametrically by Lemaitre \cite{1} in terms of the Weierstrass elliptic phi-function, sigma function, and zeta function.  Special case solutions were also obtained by R. Tolman, H. Bondi, B. Datta, as referenced in the paper of G. Omer \cite{2}, for example.

More generally the present paper, which is a continuation of work initiated in \cite{3}, concerns a parametric solution of the nonlinear differential equation
\begin{equation}\dot{y}(t)^2=\frac{ f(y(t))}{y(t)^{2n}},\end{equation}
with applications, where $f(x)=a_0x^4+4a_1x^3+6a_2x^2+4a_3x+a_4$ is a quartic polynomial and $n\geq 0$ is a fixed whole number.  Even the case $n=1$ is of some special interest as it includes equation (1.2), for example.  Namely, for $n=1$ we write (1.3) as 
\begin{equation}
\dot{y}(t)^2=By(t)^2+Ey(t)-K+\frac{A}{y(t)}+\frac{D}{y(t)^2},
\end{equation}
which is (1.2) for $E=0, D=0$.  The case $n=2$, for example, is of relevance in a discussion of  a Bianchi V cosmological model -- as we shall see.  Interestingly enough, similar to Lemaitre's solution of (1.2), a solution to (1.3) is also expressed in terms of Weierstrass' phi, sigma, and zeta functions -- as we shall also see.  Moreover, following Lemaitre, we also express our solution in terms of \emph{theta functions} (introduced by C. Jacobi in 1829) since the series for these functions are known to converge quite rapidly and are therefore amenable to practical calculations.

An additional impetus for the present work is the nice paper of G. Krani-\\
otis and S. Whitehouse \cite{4} which presents the most general exact solution of inhomogeneous relativistic cosmology, based on the Szekeres-Szafron family of metrics \cite{5, 6} (also see \cite{7})
\begin{equation}ds^2=dt^2-e^{2\beta(t,x,y,r)}\left(  dx^2+dy^2  \right) - e^{2\alpha(t,x,y,r)}dr^2,\end{equation}
where again pressureless matter is assumed, and the condition $\frac{\partial \beta}{\partial r}\neq 0$ is imposed.  The other case $\frac{\partial \beta}{\partial r}=0$ provides for a family that generalizes Friedmann and Kantowski-Sachs models \cite{8}.  The functions $\alpha, \beta$, by way of the Einstein field equations, assume the form
\begin{eqnarray}
e^{\beta(t,x,y,r)}
&=&\Phi(t,r)e^{\nu(r,x,y)},\nonumber\\
&&\\
e^{\alpha(t,x,y,r)}&=&h(r)\Phi(t,r)\frac{\partial \beta}{\partial r}(t,x,y,r),\nonumber
\end{eqnarray}
where $\nu(r,x,y)$ has the form 
\begin{equation}
e^{-\nu(r,x,y)}=A(r)(x^2+y^2)+2B(r)x+2C(r)y+D(r)
\end{equation}
and where $\Phi(t,r)$ satisfies
\begin{equation}
\Phi(t,r)\frac{\partial \beta}{\partial r}(t,x,y,r)=\frac{\partial \Phi}{\partial r}(t,r)+\Phi(t,r)\frac{\partial \nu}{\partial r}(r,x,y),
\nonumber\end{equation}
\begin{equation}
4\left[   A(r)D(r)-B(r)^2-C(r)^2 \right] = h(r)^{-2}+K(r),
\end{equation}
\begin{equation}
\left(  \frac{\partial \Phi}{\partial t} \right)^2(t,r)=-K(r)+\frac{2M(r)}{\Phi(t,r)}+\frac{\Lambda}{3}\Phi(t,r)^2.\nonumber
\end{equation}
Thus the third equation in (1.8), which (mathematically) is equation (1.2) is (again) equation (1.4) with $E=0, D=0$.

Solutions of (1.8) (in both cases $\frac{\partial \beta}{\partial r}\neq 0$ and $\frac{\partial \beta}{\partial r}=0$) were also found by J. Barrow and J. Stein-Schabes \cite{9}, for example.  Also compare the earlier paper \cite{10} of G. Covarrubias.  The 1997 text \cite{7} by A. Krasinski can be consulted for a rather comprehensive analysis of inhomogeneous cosmological models.  This is supplemented by the 2011 \emph{Review Article} by K. Bolejko, M.-No\"elle Celerier, and A. Krasinski \cite{11} -- the main thrust of which is a description of various cosmological observations that are explainable by the models of Lemaitre-Tolman and Szekeres.

Part of our interest here is to extend portions of the Kraniotis-Whitehouse (K-W) paper \cite{4}.  For example:  (i) We define and compute Hubble and deceleration parameters for our solution of (1.4); the results are expressed in terms of the Weierstrass phi-function $\wp(w)$ (ii) Whereas $D$ in (1.4) is zero in the K-W paper, we consider the case $D\neq 0$, which amounts to a constant electric charge contribution to the third dynamical equation in (1.8).  Here we find in the case of a vanishing cosmological constant a new (non-parametrized) solution in terms of J. Lambert's W-function.  In the literature this $W$-function is also called the \emph{product logarithm} or the \emph{Omega function}.  Namely, it is an inverse function, given a particular branch, of the function $f(w)=we^w$ (iii) We consider briefly the zeros of $\wp(w)$ in the general context of equation (1.4).  In \cite{4} an expression is derived for the Hubble parameter in the de-Sitter phase (the inflationary scenario) at such a zero, where the zeros are described there by the 1982 \emph{Eichler-Zagier formula} \cite{12}.   There is, on the other hand, a quite different, newer formula due to Duke-Imamoglu \cite{13} which we present (for the record) as it may be less familiar to Physicists (iv) We consider an \emph{automorphic} (or \emph{modular}) \emph{property} of our solution.  Such a (deep) property is referred to (but not explicated) in the paper \cite{14} of E. Abdella and L. A. Correa-Borbonet, in the special case of a Friedmann universe -- in connection with conformal field theory on a torus and J. Cardy's entropy formula -- the torus being constructed from a ratio of the two periods of the Weierstrass function $\wp(w)$.  These authors also take the K-W paper as impetus for their work.  

Apart from the K-W paper and an application of the solution of (1.3) to Bianchi V cosmology, we find in addition (v) a family of new solutions for a Bianchi IX model in Lyra geometry.  This extends some initial work of G. Bag, B. Bhui, S. Das, and F. Rahaman \cite{15}.  (vi) Some details are provided that indicate why the differential equation for the \emph{second moment} $I_2(t)$ of a wave function of the Gross-Pitaevskii equation is also an example of (1.4), with $A=B=0$.  Thus we can compute this moment (for a stiff fluid model), which is essential for the dynamic correspondence set up in \cite{16, 17} between Friedmann-Lemaitre-Robertson-Walker (FLRW) and/or Bianchi I cosmology and Bose-Einstein condensates -- a correspondence which one may regard as between cosmology and condensed matter -- or (very generally) between a gravitational and a non-gravitational system.

Various examples are presented, together with $3$ appendices, to assist in the navigation between theory and application.

As is well-known, the concept of entropy in the arena of quantum statistical mechanics was developed by John von Neumann.  One of his co-authors, and a staff member of his electronic computer project at the Institute of Advanced Studies at Princeton during the year 1952-1953 was A. Edward (Eddie) Nussbaum, to whom this paper is dedicated.  Professor Nussbaum was one of the most influential teachers and a mentor to the second named author -- who expresses herewith his manifold appreciation for that influence and inspiration over the course of many years.

\section{The Solution of (1.3) in Terms of Theta Functions}

\indent \indent  Associated with the polynomial $f(x)$ in (1.3) are the \emph{invariants} $g_2, g_3$ and the \emph{discriminant} $\Delta$ given by
\begin{eqnarray}
g_2&\stackrel{def.}{=}&a_0a_4-4a_1a_3+3a_2^2,\nonumber\\
g_3&\stackrel{def.}{=}&a_0a_2a_4+2a_1a_2a_3-a_2^3-a_0a_3^2-a_1^2a_4=\left|
\begin{array}{ccc}
a_0&a_1&a_2\\
a_1&a_2&a_3\\
a_2&a_3&a_4
\end{array}
\right|,\nonumber\\
\Delta&\stackrel{def.}{=}& g_2^3-27g_3^2.
\end{eqnarray}
Let $e_1, e_2, e_3$ denote the roots of $4x^3-g_2x-g_3=0$ and fix a root $x_0$ of $f(x)=0$, which we assume is \emph{non-repeated}.  If the leading coefficient $a_0$ of $f(x)$ is indeed \emph{non-zero}, then $x_0$ is non-repeated $\Leftrightarrow \Delta \neq 0$, as shown in chapter 3 of \cite{18}, for example.  In \cite{3} we have derived the following parametric solution of equation (1.3), where we refer to Appendix A following section 4 of this paper for a review of the definitions of the classical Weierstrass functions $\wp(w)=\wp(w; \omega_1, \omega_2), \sigma(w)$, and $\zeta(w)$:
\begin{eqnarray}
y&=& x_0+\frac{f'(x_0)}{4\left[  \wp(w+c)-f''(x_0)/24 \right]},\\
t&=&\displaystyle\int \left(  x_0 + \frac{f'(x_0)}{4 \left[  \wp(w+c)-f''(x_0)/24\right] } \right) ^ndw+\delta,
\end{eqnarray}
for integration constants $c,\delta$.  Moreover, for a choice of any $w_0$ with $\wp(w_0)=f''(x_0)/24$, $\wp(w_0)\neq e_1, e_2, e_3$ we have for $n=1,2$ respectively in (2.3)
\begin{eqnarray}
t&=& x_0w+\frac{f'(x_0)}{4\wp'(w_0)} \left[  \log \frac{\sigma(w+c-w_0)}{\sigma(w+c+w_0)} +2(w+c)\zeta(w_0) \right] + \delta, \\
t&=& x_0^2w+\left[  -\frac{x_0 f'(x_0)}{2\wp'(w_0)}+ \frac{ f'(x_0)^2 \wp''(w_0)}{16\wp'(w_0)^3 }\right] \log \frac{\sigma(w+c+w_0)}{\sigma(w+c-w_0)}\nonumber\\
&& -\frac{f'(x_0)^2}{16\wp'(w_0)^2} \left[  \zeta(w+c+w_0)+\zeta(w+c-w_0) \right] \\
&& +(w+c) \left(  \frac{x_0 f'(x_0)}{\wp'(w_0)} \zeta(w_0) - \frac{f'(x_0)^2}{16} \left[  \frac{2\wp(w_0)}{\wp'(w_0)^2} + \frac{2\wp''(w_0)\zeta(w_0)}{\wp'(w_0)^3 } \right] \right)+\delta,\nonumber
\end{eqnarray}
by formulas 1037.06, 1037.11, respectively, in \cite{19}.  In case $n=0$, one has the (non-parametrized) Biermann-Weierstrass solution \cite{20, 21, 22}
\begin{equation}
y(t)=y(0)+ \frac{  
\left[ f(y(0))^{1/2}\wp'(t)+\frac{f'(y(0))}{2}\left(  \wp(t)-\frac{f''(y(0))}{24}\right)  + \frac{ f(y(0))f''''(y(0))}{24} \right]}
{2 \left[ \wp(t)-\frac{f''(y(0))}{24}\right]^2 - \frac{f(y(0))f''''(y(0))}{48}}
\end{equation}
of (1.3).  Here the periods $\omega_1, \omega_2$ of $\wp(t)$ are constructed from the invariants $g_2, g_3$ of $f(x)$ in (2.1), according to Appendix A.  Formula (2.6) simplifies quite a bit in case $y(0)$ is a root of $f(x)$, which is the case if, for example, the boundary condition $\dot{y}(0)=0$ is imposed:
\begin{equation}
y(t)=y(0)+ \frac{f'(y(0))}{4 \left[  \wp(t)-\frac{f''(y(0))}{24} \right] },
\end{equation}
which is reminiscent of equation (2.2).  One can replace $y(0)$ in (2.7) by any non-repeated root $x_0$ of $f(x)$ in fact.  We specifically choose the order of the roots $e_j$, $j=1,2,3$, by setting
$e_j = \wp(\omega_j/2; \omega_1, \omega_2)$, where $\omega_3 \stackrel{def.}{=} \omega_1 + \omega_2$.

We proceed now to a theta function expression of the solution -- the importance of which was indicated in the introduction.  There is, unfortunately, no uniformity in the definitions/notation of theta functions as indeed considerable variations exist across various texts.  Here we shall employ, specifically, the definitions and notation of the Chandrasekharan text \cite{18}, together with certain results therein:  For a complex number $z\in\mathds{C}$ and $\tau\in$ the upper $\frac{1}{2}$-plane $\Pi^{+}$ (i.e. $Im\tau>0$), and $q\stackrel{def.}{=}e^{\pi i \tau}$
\begin{eqnarray}
\theta(z, \tau) & \stackrel{def.}{=}& 2 \displaystyle\sum_{n=0}^\infty (-1)^n q^{(n+\frac{1}{2})^2} \sin((2n+1)\pi z),\nonumber\\
\theta_1(z,\tau)&\stackrel{def.}{=}&2\displaystyle\sum_{n=0}^\infty q^{(n+\frac{1}{2})^2}\cos((2n+1)\pi z),\nonumber\\
\\
\theta_2(z, \tau)&\stackrel{def.}{=}&1+2\displaystyle\sum_{n=1}^\infty (-1)^nq^{n^2}\cos(2\pi nz),\nonumber\\
\theta_3(z,\tau)&\stackrel{def.}{=}&1+2\displaystyle\sum_{n=1}^\infty q^{n^2}\cos(2\pi nz).\nonumber
\end{eqnarray}
For $\tau$ fixed, these are entire functions of $z$.  Obviously $\theta(z,\tau)$ is an odd function of $z$ and the $\theta_j(z,\tau)$, $j=1,2,3,$ are even functions of $z$.  

For the choice $\tau=\omega_2/\omega_1\in\Pi^{+}$, the theta function expression of $\wp(w)=\wp(w; \omega_1, \omega_2)$ is given by
\begin{equation}
\wp(w)=e_j+\frac{1}{w_1^2}\left[ \frac{\theta'(0,\tau)}{\theta_j(0,\tau)}\right]^2 \left[ \frac{\theta_j\left( \frac{w}{\omega_1}, \tau\right)}{\theta\left( \frac{w}{\omega_1},\tau\right)}\right]^2
\end{equation} 
for $j=1,2,3,$ where $\theta'(z,\tau)$ denotes partial differentiation with respect to $z$.  Similarly there are theta function expressions of the sigma and zeta functions given by 
\begin{eqnarray}
\sigma(w)&=& \frac{\omega_1}{\theta'(0,\tau)}e^{\eta_1w^2/\omega_1}\theta\left( \frac{w}{\omega_1}, \tau\right),\nonumber\\
\\
\zeta(w)&\stackrel{def.}{=} &\frac{\sigma'(w)}{\sigma(w)}\stackrel{\therefore}{=}\frac{2\eta_1 w}{\omega_1}+\frac{1}{\omega_1}\frac{\theta'\left(  \frac{w}{\omega_1}, \tau\right)}{\theta\left( \frac{w}{\omega_1}, \tau\right) },\nonumber
\end{eqnarray}
for $\tau\stackrel{def.}{=} \omega_2/\omega_1$, $\eta_1\stackrel{def.}{=} \zeta\left( \omega_1/2\right).$  In particular
\begin{equation}
\frac{\sigma(u-w_0)}{\sigma(u+w_0)}=e^{-4\eta_1 w_0 u/\omega_1} \frac{\theta\left(  \frac{u-w_0}{\omega_1}, \tau \right) }{\theta\left(  \frac{u+w_0}{\omega_1}, \tau \right)},\end{equation}
which with (2.10) gives
\begin{eqnarray}
&&\log \frac{\sigma(w+c-w_0)}{\sigma(w+c+w_0)}+2(w+c)\zeta(w_0)=\nonumber\\
\\
&& \log \frac{\theta\left( \frac{w+c-w_0}{\omega_1},\tau\right)}{\theta\left(\frac{w+c+w_0}{\omega_1},\tau\right)}+ \frac{2(w+c)\theta'\left(  \frac{w_0}{\omega_1}, \tau\right)}{\omega_1 \theta\left(  \frac{w_0}{\omega_1}, \tau\right)},\nonumber
\end{eqnarray}
where in (2.12) the constant $\eta_1$ no longer appears.  Strictly speaking, in the argument for (2.12) one has $\log(z_1z_2)=\log z_1 +\log z_2 + 2\pi i N$, $\log e^z=z$, for $N\in\left\{ -1, 0, 1\right\}, -\pi<Im(z) \leq \pi$, for the principal branch of the logarithm.  However we absorb the $2\pi i N$ into the integration constant $\delta$ in (2.3).  That is, in summary we can express the solution of equation (1.4) parametrically by $y$ and $t$ in (2.2) and (2.4), or equivalently (using (2.9), (2.12)) by
\begin{eqnarray}
y&=&x_0 + \frac{f'(x_0)}{4}\left\{  e_j-\frac{f''(x_0)}{24}+\frac{1}{\omega_1^2} \left[ \frac{\theta'(0, \tau)}{\theta_j(0, \tau)} \right]^2 \left[  \frac{\theta_j\left(  \frac{w+c}{\omega_1}, \tau\right) }{\theta\left( \frac{w+c}{\omega_1}, \tau \right) } \right]^2 \right\}^{-1},  \  \ \  \ \ \ \ \\
t&=& x_0w+\frac{f'(x_0)}{4\wp'(w_0)} \left[  \log \frac{ \theta\left(  \frac{w+c-w_0}{\omega_1}, \tau \right)}{  \theta\left(  \frac{w+c+w_0}{\omega_1}, \tau \right)} + \frac{2(w+c) \theta' \left( \frac{w_0}{\omega_1}, \tau\right) }{\omega_1 \theta\left(  \frac{w_0}{\omega_1}, \tau\right)} \right] + \delta,
\end{eqnarray}
for $j=1,2,3$, $\tau=\omega_2/\omega_1$, $x_0=$ a non-repeated root of $f(x)=0$, and for the theta functions defined in (2.8); the $e_j$, $w_0$ are defined following (2.1) and (2.3).  Here for $n=1$ we have $f(x)=Bx^4+Ex^3-Kx^2+Ax+D$, in accordance with the notation in (1.4). 

Equation (2.13) for $y$ (or equivalently equation (2.2)) holds independently of \emph{whatever value} $n$ assumes.  Only the expression for $t$ (as a function of $w$) depends on $n$, by (2.3).  For $n=2$, for example, $t$ is already given by formula (2.5), which we also express in terms of theta functions as follows:
\begin{eqnarray}
\hspace{-.15in}t&=&x_0^2w+ \left[  \frac{-x_0f'(x_0)}{2\wp'(w_0)} + \frac{f'(x_0)^2\wp''(w_0)}{16 \wp'(w_0)^3} \right] \left[ 4\eta_1 \frac{w_0}{\omega_1}(w+c) 
+ \log \frac{ \theta \left(  \frac{w+c+w_0}{\omega_1}, \tau \right) }{  \theta \left(  \frac{w+c-w_0}{\omega_1}, \tau \right) }\right]\nonumber\\
\\
&& - \frac{f'(x_0)^2}{16\omega_1 \wp'(w_0)^2} \left[  4\eta_1(w+c) + 
 \frac{ \theta' \left(  \frac{w+c+w_0}{\omega_1}, \tau \right) }{  \theta \left(  \frac{w+c+w_0}{\omega_1}, \tau \right) } +  \frac{ \theta' \left(  \frac{w+c-w_0}{\omega_1}, \tau \right) }{  \theta \left(  \frac{w+c-w_0}{\omega_1}, \tau \right) }\right]\nonumber\\
&& +(w+c)\left(  \frac{x_0 f'(x_0)}{\wp ' (w_0)} \zeta(w_0) - \frac{f'(x_0)^2}{16} \left[  \frac{2\wp(w_0)}{\wp'(w_0)^2} + \frac{2\wp''(w_0) \zeta(w_0)}{\wp'(w_0)^3} \right] \right) +\delta\nonumber,
 \end{eqnarray}
by (2.10) and (2.11), where (again) $\eta_1 \stackrel{def.}{=}\zeta(\omega_1/2)$.

Although the cases $n=0, 1, 2$ treated here are sufficient for the physical applications we have in mind, one can in principle carry out the integration in (2.3) for $n\geq 3$ by using the recursion formula 1037.12 on page 312 of \cite{19}.

Naturally associated with the solution (2.2), (2.4) of equation (1.4) are the \emph{Hubble} and \emph{deceleration parameters}
\begin{equation}
H(t)\stackrel{def.}{=}\frac{ \dot{y}(t)}{y(t)},  \qquad q(t)\stackrel{def.}{=}  - \frac{\ddot{y}(t)y(t)}{\dot{y}(t)},
\end{equation}
respectively.  Using that (2.4) was derived by integration of (2.3) -- i.e.
\begin{equation}
\frac{dt}{dw}=x_0 + \frac{f'(x_0)}{4\left[ \wp(w+c) - f''(x_0)/24\right] } 
\end{equation}
for $n=1$, one applies parametric differentiation to deduce the following formulas:
\begin{equation}
H(t)= \frac{ -4f'(x_0)\wp'(w+c)} { \left( 4x_0 \left[  \wp(w+c) - \frac{f''(x_0)}{24} \right] + f'(x_0) \right) ^2 },\nonumber
\end{equation}
\begin{eqnarray}
\hspace{-.325in}&&q(t)=\\
\hspace{-.325in}&& \left(  \frac{\wp''(w+c) \left[  \wp(w+c)-\frac{f''(x_0)}{24}\right] }{\wp'(w+c)^2 } -2 \right) \left( \frac{4x_0 \left[ \wp(w+c)-\frac{f''(x_0)}{24} \right] + f'(x_0)}{f'(x_0)} \right) +1 .\nonumber
\end{eqnarray}
Here $f'(x_0)=4Bx_0^3+3Ex_0^2-2Kx_0+A$, $f''(x_0)/24=\frac{B}{2}x_0^2 + \frac{E}{4}x_0 - \frac{K}{12}$.  These formulas simplify in case $x_0=0$ is a root of $f(x)=0$ -- say (for example) $D=0$ in (1.4), the case considered in \cite{4}:
\begin{eqnarray}
H(t)&=& \frac{-4\wp'(w+c)}{f'(0)}\nonumber\\
&&\\
q(t)&=& \frac{ \wp''(w+c) \left[ \wp(w+c)- f''(0)/24 \right]}{\wp ' (w+c)^2} - 1;\nonumber
\end{eqnarray}
compare formulas (48) and (50) in \cite{4}.  Since $\wp(z)$ satisfies the differential equation
\begin{equation}
\wp'(z)^2=4\wp(z)^3-g_2\wp(z)-g_3,
\end{equation}
one can use that $\wp''(w+c)=6\wp(w+c)^2-g_2/2$ in formulas (2.18), (2.19).  Note that a slight misprint occurs in formula (46) of \cite{4} for $\dot{R}(t)$.  Namely, a minus sign is needed.  Similarly formula (65) should read $\wp'(w_0)=-i\sqrt{g_3}$.

We have assumed so far that the root $x_0$ was non-repeated.  For repeated roots, solutions of (1.3) in fact are generally easier to obtain and they usually involve elementary, non-elliptic functions.  As an example, take $f(x)=-4x^4+2x^2+x/\sqrt{2}+1/16$ which has $x_0\stackrel{def.}{=}-\sqrt{2}/4$ as a repeated root:  $(x-x_0)^2$ is a factor of $f(x)$.  Equation (1.3) (with $n=0$) is $u'(x)^2=-4u(x)^4+2u(x)^2+u(x)/\sqrt{2}+1/16$, which has elementary, non-elliptic solutions 
\begin{equation}u(x)=u_{\mp}(x; \delta)\stackrel{def.}{=} \frac{  \mp \sin(x-\delta)}{ 2\sqrt{2}\left[  \sqrt{2}\pm \sin(x-\delta) \right] },\end{equation}
as discussed in \cite{3, 23}, for example.  One can ``deform" the solution $u_+(x;0)$ to obtain a solution $u(x,t)$ of the \emph{modified Novikov-Veselov equation}
\begin{equation}
u_t=u_{xxx}+24u^2u_x.
\end{equation}
Namely, one takes 
\begin{equation}
u(x,t)\stackrel{def.}{=}u_+(x+2t;0)\stackrel{def.}{=} \frac{\sin(x+2t)}{2\sqrt{2}\left[  \sqrt{2}- \sin(x+2t)\right] }.
\end{equation}

Note that a solution of the differential equation
\begin{eqnarray}
\dot{y}_1(t)^2&=&y_1(t)^2\left[  B_1+E_1y_1(t)^m-K_1y_1(t)^{2m}+A_1y_1(t)^{3m}+D_1y_1(t)^{4m}\right],\nonumber\\
&&
\end{eqnarray}
where $m\neq 0$ is a fixed whole number, can be obtained from a solution $y(t)$ of equation (1.4).  For this we set $y(t)=y_1(t)^{-m}$.  Then for $B=m^2B_1$, $E=m^2E_1$, $K=m^2K_1$, $A=m^2A_1$, and $D=m^2D_1$, equation (2.24) is transformed to equation (1.4).

\section{A Modular and Elliptic Property of the Solution of (1.3)}

\indent \indent The phi-function is homogeneous of degree $-2$ in the sense that for any $\lambda\in\mathds{C}-\left\{ 0\right\}$, $\wp(\lambda w; \lambda \omega_1, \lambda \omega_2) = \lambda^{-2}\wp(w; \omega_1, \omega_2).$  In particular for $\lambda=\omega_1^{-1}$ and $\tau=\frac{\omega_2}{\omega_1}\in\Pi^+$ (as in (2.10)) $\wp(w; \omega_1, \omega_2)=\omega_1^{-2}\wp\left(\frac{w}{\omega_1}; 1; \tau\right)$, which means that the phi-function is determined by the special phi-function $\wp(w; 1; \tau)$ that we focus on here, and which we denote by $\wp(w; \tau)$ for an arbitrary $\tau\in\Pi^+$.  $\wp(w; \tau)$ satisfies the wonderful \emph{modular property} \cite{24}
\begin{equation}
\wp\left(  \frac{w}{c\tau + d}; \frac{a\tau+b}{c\tau+d} \right) = (c\tau+d)^2\wp(w; \tau) 
\end{equation}
for $\gamma=\left[  \begin{array}{cc}a&b\\ c&d \end{array}\right] \in SL\left(  2, \mathds{Z} \right)$; i.e. $a,b,c,d\in\mathds{Z}$ and $det\gamma = 1$.  The \emph{elliptic } (or \emph{quasi-periodicity}) \emph{property}
\begin{equation}
\wp(w+m\tau+n; \tau)=\wp(w; \tau)
\end{equation}
is also satisfied for $m,n\in\mathds{Z}$.  The modular property is fundamental for the discussion in \cite{14} regarding solutions of the Friedmann equation (in various settings such as that of a $4$-dimensional radiation dominated universe, for example) and connections to conformal field theory (CFT) on a torus (defined by the modular parameter $\tau$), and with some linkage to the Cardy entropy formula.  We offer a bit more on this later.  Here $\Delta(1,\tau)\neq 0$ (see Appendix A) and $\tau$ depends on the cosmological constant.

We illustrate how the modular property translates to a concrete modular property of the solution of equation (1.3).  Thus we incorporate the dependence of the solution on $\tau$ and write $y(w; \tau)$ for $y$ in (2.2), where the choice of lattice in Appendix A is $\mathscr{L}=\mathscr{L}(1, \tau)$:
\begin{equation}
y(w;\tau)=x_0+ \frac{f'(x_0)}{4\left[  \wp(w; \tau)-f''(x_0)/24 \right] } .
\end{equation}
Here we have chosen the integration constant $c$ in (2.2) to be zero.  Then for $\gamma=\left[  \begin{array}{cc}a&b\\ c&d\end{array} \right]\in SL\left( 2,\mathds{Z} \right)$ with $\gamma\cdot\tau \stackrel{def.}{=} \frac{a\tau+b}{c\tau+d}$
\begin{equation}
y\left( \frac{w}{c\tau+d}; \gamma\cdot\tau \right) = x_0 + \frac{f'(x_0)}{4 \left[  \wp\left( \frac{w}{c\tau+d}; \gamma\cdot\tau\right) - \frac{f''(x_0)}{24} \right]}.
\end{equation}
The idea now is simple:  In (3.4), apply (3.1) and then use (3.3) to express $\wp(w; \tau)$ in terms of $y(w; \tau)$.  The result is 
\begin{eqnarray}
&&\hspace{-.4in} y\left(  \frac{w}{c\tau+d}; \frac{a\tau+b}{c\tau+d} \right) =\\
&&\hspace{-.1in}x_0  + \frac{f'(x_0)[  y(w; \tau)-x_0 ]}{ \frac{f''(x_0)}{6}\left[  (c\tau+d)^2-1 \right] \left[ y(w; \tau)-x_0 \right] +f'(x_0)(c\tau+d)^2}.\nonumber
\end{eqnarray}
In addition to the modular property (3.5) of the solution (3.3), equation (3.2) provides for the (less profound) elliptic property
\begin{equation}
y(w+m\tau+n; \tau)=y(w; \tau)
\end{equation}
for $m,n\in\mathds{Z}$.

If (again) $x_0=0$ is a root, our formulas simplify.  Suppose in fact, for example, that also $f''(0)=0$ which means that $K=0$ in (1.4).  Then equation (3.5) reduces to 
\begin{equation}
y\left( \frac{w}{c\tau+d}; \frac{a\tau+b}{c\tau+d}\right) = (c\tau+d)^{-2}y(w; \tau),
\end{equation}
which also follows directly from (3.1), (3.3).  The case $K=0$ is discussed in section 5 of \cite{4}, where a flat universe with non-zero cosmological constant is considered.  One can't help but note the comparison of (3.7) with (3.1).

The authors in \cite{14}, seeking new connections between the Friedmann equation and Cardy formula, set up the ``chain of connections" (using their phase on page 5):  Friedmann equation $\longrightarrow$ Weierstrass equation $\longrightarrow$ torus $\longrightarrow$ CFT partition function $Z(\tau)$ $\longrightarrow$ Cardy formula, where for $q\stackrel{def.}{=}e^{2\pi i \tau}$
\begin{equation}
Z(\tau)=trace \ q^{L_0 - c/24}=trace \ e^{-\beta H}
\end{equation}
with $c=$ a central charge, $H=L_0 - c/24$ a Hamiltonian corresponding to the Virasoro generator $L_0$, $\beta=-2\pi i\tau$ an inverse temperature, and with the Cardy entropy formula \cite{25} given by 
\begin{equation}
S=2\pi \sqrt{ \frac{c}{6} \left(  L_0 - \frac{c}{24} \right) }.
\end{equation}
Here $L_0$ also denotes the eigenvalue of $L_0$ and, for simplicity, the anti-holomorphic sector (which involves $\overline{\tau}$) is not considered.  As mentioned above $\tau$ depends on the cosmological constant $\Lambda$; it also depends on the spacetime dimension.

We are positioned to mimic this chain of connections as follows.  The Friedmann equation is replaced, more generally, by equation (1.3).  The Weierstrass equation (or solution) is given by (3.3) (and by (2.3)).  $\tau$ in (3.3) defines a corresponding complex torus $T\stackrel{def.}{=} \mathds{C}/ \mathscr{L}(1, \tau)$, which (as is well known, based on (2.20)) is an elliptic curve parametrized by the phi-function and its derivative.  The remainder of the chain is as before, given by (3.8), (3.9). 

John Cardy's great insight was that the key property of \emph{modular invariance} of CFT partition functions was deeply consequential.  In particular it lead to his asymptotic density of states formula and hence to his entropy formula -- (3.9) being a special case of the general formula that also incorporates the non-holomorphic sector.

If we write (3.1), (3.2) as 
\begin{eqnarray}
\wp\left( \frac{w}{c\tau+d}; \frac{a\tau+b}{c\tau+d} \right) &=& e^{2\pi i j  \frac{cw^2}{c\tau+d}} (c\tau+d)^k \wp(w; \tau),\nonumber\\
&&\\
\wp(w+m\tau+n; \tau) &=& e^{-2\pi ij (m^2 \tau+2mw)} \wp(w; \tau)\nonumber
\end{eqnarray}
for $j,k\in\mathds{Z}$ (namely for $j=0, k=2$), we see that $\wp(w; \tau)$ is a (meromorphic) \emph{Jacobi form of index} $j=0$ and \emph{weight} $k=2$ \cite{24}.  Similarly, given (3.6), (3.7), we can regard $y(w; \tau)$ as a Jacobi form of index $0$ and weight $-2$.  Strictly speaking, a growth condition at infinity is also imposed in the definition of a Jacobi form (which we shall not have a concern with here) -- a condition analogous to that imposed in the definition of a modular form \cite{26}.  The \emph{elliptic genus} of an $N=(2,2)$ \emph{Super} CFT with central charge $c=6j$ is, for example, a Jacobi form of index $j$ and weight $k=0$.

\newpage

\section{Some Applications}

\indent \indent  We present some examples to provide further clarity and applications of the formulas in section 2.\\

\noindent {\bf Example 1.}  Among various modifications of Einstein's GR was one introduced by G. Lyra in 1951 in which a gauge function was introduced to affect Riemannian geometry that rendered it a generalization of Weyl's 1918 geometry.  More specifically, the purpose of the gauge function was to remove the non-integrability of vector length under parallel transport.  Many cosmological and string models based on Lyra geometry have been studied.  In particular the Bianchi IX model in this geometry was studied in \cite{15}.  Here, for the metric
\begin{eqnarray}
ds^2&=&-dt^2+a(t)^2dx^2+b(t)^2dy^2+\left[  b(t)^2\sin^2y+a(t)^2 \cos^2 y\right] dz^2\nonumber\\
&&\qquad -2a(t)^2 \cos y dxdz
\end{eqnarray}
and a constant flat potential $V(\phi)=2\lambda$, the modified Einstein equations yield the relation $\dot\phi=\phi_0/ab^2$ for an integration constant $\phi_0$, and the assumption $a=b^n$ leads moreover to the first integral

\begin{equation}
\dot{b}^2=\frac{1}{n^2-1}-\frac{ b^{2n-2}}{  2(n^2-n)}  +D_1b^{-2n-2},
\end{equation}
for $n\neq 0, \pm 1$ and an integration constant $D_1$.  The authors in \cite{15} found solutions of (4.2) only for $D_1=0$, $n=2, \frac{1}{2}, \frac{3}{2}$, and $\frac{3}{4}$.  A solution for $n=2$, $D_1\neq 0$ (say $D_1=\frac{7}{12}$) was found in \cite{3}.  We indicate here how to construct an infinite family of solutions with $\Delta\neq 0$ corresponding to
\begin{equation}
D_1=D_1(\lambda)\stackrel{def.}{=} \frac{3\lambda^4+4\lambda^3}{12}, \lambda\neq 0, -1, -\frac{4}{3}, \frac{1 \pm \sqrt{2}i}{3},
\end{equation} 
where for $\lambda=1$ we obtain the solution just mentioned, corresponding to $D_1=\frac{7}{12}.$

For $n=2$, $a=b^2$ and equation (4.2) is transformed to the equation
\begin{equation}
\dot{a}(t)^2=-a(t)^2+\frac{4}{3}a(t)+\frac{4D_1}{a(t)^2},
\end{equation}
which is equation (1.4) for $B=-1, E=\frac{4}{3}$, $K=0$, $A=0$, $D=4D_1$.  From (2.1), $g_2=-4D_1$, $g_3=-4D_1/9=g_2/9$, and $\Delta=-16D_1^2\left( 4D_1 + \frac{1}{3}\right).$  $\Delta=0$  $\Rightarrow$ $D_1=0, -\frac{1}{12}$, which for $D_1=D_1(\lambda)$ in (4.3) means that $\lambda=0, -\frac{4}{3}$ and $3\lambda^4+4\lambda^3=-1$.  The latter equation is $0=3\lambda^4+4\lambda^3+1=(\lambda+1)^2(3\lambda^2-2\lambda+1),$ which gives $\lambda=-1, (1\pm \sqrt{2}i)/3$.  That is, by (4.3), $\Delta\neq 0$ for $D_1=D_1(\lambda)$ so by the remarks following (2.1) $f(x)=-x^4+\frac{4}{3}x^3+4D_1(\lambda)=0$ has no repeated roots, where we note that $x_0=-\lambda$ is indeed a root.  As $f'(-\lambda)=4\lambda^2(\lambda+1)$ and $f''(-\lambda)/24=-\lambda(3\lambda+2)/6$, we obtain the parametric solutions
\begin{eqnarray}
a&=&-\lambda+\frac{\lambda^2(\lambda+1)}{\wp(w+c)+\lambda(3\lambda+2)/6},\nonumber\\
&&\\
t&=& -\lambda w + \frac{\lambda^2(\lambda+1)}{\wp'(w_0)} \left[  \log \frac{ \sigma(w+c-w_0)}{\sigma(w+c+w_0)}+2(w+c)\zeta(w_0)\right] + \delta\nonumber
\end{eqnarray}
of equation (4.4), by formulas (2.2), (2.4), for $\lambda\neq 0, -1, -\frac{4}{3}, (1\pm\sqrt{2} i)/3$.  Then in the metric (4.1), $b(t)^2 = a(t)$.\\

\noindent {\bf Example 2.}  Consider next, briefly, the metric
\begin{equation}
ds^2 = -dt^2+X(t)^2dx^2 + e^{2bx}\left[  Y(t)^2dy^2+Z(t)^2 dz^2 \right]
\end{equation}
for a Bianchi $V$ anisotropic cosmological model, where $b\neq 0$ is a constant.  $R(t)\stackrel{def.}{=} \left[ X(t)Y(t)Z(t) \right]^{1/3}$, as a consequence of the Einstein equations, satisfies a differential equation 
\begin{equation}
\dot{R}(t)^2 = f(R(t))/R(t)^4
\end{equation}
for a suitable $4^{th}$ degree polynomial $f(x)$ with leading coefficient $b^2$.  A key, useful observation here is that the quantity/function
\begin{eqnarray}
\hspace{-.325in}&&D_0(t) \stackrel{def.}{=}\\
\hspace{-.325in}&& R(t)^2 \left[  \left( \frac{\dot{X}(t)}{X(t)} \right)^2 + \left( \frac{\dot{Y}(t)}{Y(t)} \right)^2 + \left( \frac{\dot{Z}(t)}{Z(t)} \right)^2 - \frac{\dot{X}(t)\dot{Y}(t)}{X(t)Y(t)} - \frac{\dot{X}(t)\dot{Z}(t)}{X(t)Z(t)} - \frac{\dot{Y}(t)\dot{Z}(t)}{Y(t)Z(t)} \right] \nonumber
\end{eqnarray}
is actually a constant, independent of $t$, which we may therefore denote by $D_0$.  $D_0$ is used, in fact, to construct the \emph{constant term} of $f(x)$.  Other coefficients of $f(x)$ involve radiation and matter constants, and the gravitational constant, where we assume that the energy momentum tensor is given by that of a perfect fluid.  For full details of these remarks and further related information the reader can consult \cite{3, 27}.

Since equation (4.7) assumes the form (1.3) with $n=2$, $R(t)$ is given parametrically by $y,t$ in equations (2.2), (2.5) -- or equivalently by (2.13), (2.15).  On the other hand, the ``scale factors"  $X(t), Y(t), Z(t)$ can be determined from $R(t)$ by the formulas in \cite{3, 27}, and thus the metric (4.6) is also determined by $R(t)$.\\

\noindent {\bf Example 3.}  The influence of an electro-magnetic field on the collapse of dust is considered briefly by Krasinski in section 7 of \cite{28}.  A more extended discussion and list of references is found in his text \cite{7}; see section 2.14, for example.  Here, again with spherical symmetry assumed, the third equation in (1.8) (or, equivalently, the Lemaitre equation (1.2)) is supplemented by a term $-\left(  Q/\Phi(t,r) \right) ^2$, where $Q$ is a constant electric charge:
\begin{equation}
\dot{\Phi}(t,r)^2 = -K(r)+ \frac{2M(r)}{\Phi(t,r)}+\frac{\Lambda}{3}\Phi(t,r)^2 - \frac{Q^2}{\Phi(t,r)^2}.
\end{equation}
It will turn out (see (2.1)) that 
\begin{equation}\nonumber
g_2 = -\frac{\Lambda}{3}Q^2 +\frac{K(r)^2}{12}, \quad g_3 = \frac{\Lambda Q^2 K(r)}{18} + \frac{K(r)^3}{216} - \frac{\Lambda M(r)^2}{12},
\end{equation}
\begin{equation}
\end{equation}
\begin{eqnarray}
\Delta & = & \Lambda \left[  - \frac{\Lambda^2 Q^6}{27} - \frac{\Lambda Q^4 K(r)^2}{18} - \frac{Q^2 K(r)^4}{48} + \frac{\Lambda Q^2 K(r) M(r)^2 }{4} \right.\nonumber\\
&& \left.  + \frac{K(r)^3 M(r)^2}{48} - \frac{3\Lambda M(r)^4}{16} \right]. \nonumber
\end{eqnarray}
In particular $\Delta=0$ when the cosmological constant $\Lambda$ vanishes, in which case one expects non-elliptic solutions of (4.9).  Indeed for $\Lambda=0$, Shikin \cite{29} has found parametric solutions of (4.9) in terms of elementary, non-elliptic functions.

On the other hand, however, since (4.9) is a special case of (1.4) (with $B=\frac{\Lambda}{3}$, $E=0$, $K=K(r)$, $A=2M(r)$, $D=-Q^2$, $f(x)=\frac{\Lambda}{3}x^4 - K(r)x^2 +2M(r)x - Q^2,$ from which one computes (4.10)) we have that its general parametric solution (for arbitrary $\Lambda$) is given already by equations (2.2), (2.4):
\begin{eqnarray}
\Phi &=& x_0 + \frac{  \frac{\Lambda}{3} x_0^3 - \frac{K(r)}{2}x_0 + \frac{M(r)}{2}}{\wp(w+c) + \frac{K(r)-2\Lambda x_0^2}{12}}, \nonumber\\
&&\\
t &=& x_0 w + \frac{\left[ \frac{\Lambda}{3}x_0^3 - \frac{K(r)x_0}{2} + \frac{ M(r) }{2} \right]}{\wp'(w_0)} \left[  \log \frac{\sigma(w+c-w_0)}{\sigma(w+c+w_0)} \right.\nonumber\\
&& \left. + 2(w+c)\zeta(w_0) \right] +\delta,\nonumber
\end{eqnarray}
where now the integration constants $c=c(r)$, $\delta=\delta(r)$ depend on $r$; throughout it is best to consider $r$ fixed.  Also, as usual, $w_0$ is a choice such that $\wp(w_0) = f''(x_0)/24 = (2\Lambda x_0^2 - K(r))/12 \neq $ the roots $e_1, e_2, e_3$ of $4x^3-g_2x-g_3 = 0$.  In general (unless $Q=0$, as in \cite{4}) the non-repeated root $x_0$ of $f(x)=0$ in (4.11) is \emph{non-zero}, which means (in particular) that the formulas in (2.18) for the Hubble and deceleration parameters do not simplify as in (2.19).  $x_0$ of course also depends on $r$.  Formulas (2.13), (2.14) provide, in addition, theta function expressions of the solution (4.11).  Although $Q$ does not appear explicitly in (4.11), the solution there is indeed $Q-$dependent as $x_0$ also depends on $Q$, and the invariants $g_2, g_3$ depend on $Q$ as well by (4.10).

Section 5.2.2 of \cite{4} contains the intriguing remark that interesting physics and mathematics arise when equation (2.20) is solved at the zeros $z_0$ of $\wp(z)$:
\begin{equation}
\wp'(z_0)^2 = -g_3. 
\end{equation}
Appendix C contains a description of these zeros (that supplements that given in \cite{4}), where (as in section 3) the choice of lattice $\mathscr{L}(1, \tau)$ is made for $\tau\in\Pi^+$ that we shall also employ here.  Particularly in section 5.2.2, a Euclidean universe with $g_2=0, g_3\neq 0, \Delta <0$ is considered.  Given the presence of the electric charge $Q$, we can slightly extend part of the discourse there as follows, where for convenience we write $K, M$ for $K(r), M(r)$.  Note first that for $g_2=0, K^2 = 4\Lambda Q^2 \Rightarrow K^3 = 4\Lambda K Q^2 \Rightarrow $
\begin{equation}
g_3 = \frac{\Lambda Q^2 K}{18} + \frac{ 4\Lambda KQ^2}{216} - \frac{\Lambda M^2}{12} = - \frac{\Lambda}{3} \left( \frac{M^2}{4} - \frac{2 K Q^2}{9} \right)
\end{equation}
by (4.10).  If we take $w=z_0- c$ in (2.18) and denote by $t_0$ the corresponding value of $t$ (where we could take $c=0$ if we wished) then using (4.12) we see that 
\begin{equation}
H(t_0)=  \frac{-4 f'(x_0) \left(  \pm \sqrt{-g_3} \right) }{ \left( -4x_0 f''(x_0)/24 + f'(x_0) \right)^2 },
\end{equation}
which by (4.13) simplifies as 
\begin{eqnarray}
H(t_0) &=& \frac{\pm 8 \left(  \frac{-2\Lambda}{3}x_0^2 +Kx_0 - M \right) }{ \left( \frac{2}{3}\Lambda x_0^3 - \frac{5}{3} Kx_0+2M \right)^2} \sqrt{  \left(  \frac{M^2}{4} - \frac{2KQ^2}{9} \right) }  \sqrt{ \frac{\Lambda}{3} }\nonumber\\
& \stackrel{\mbox{ for }M>0}{=}&  \pm \frac{ \left(  -\frac{2\Lambda x_0^2}{3M} + \frac{Kx_0}{M}-1 \right)}{\left(  \frac{\Lambda x_0^3}{3M} - \frac{5Kx_0}{6M} + 1\right)^2 } \sqrt{ 1 - \frac{8KQ^2}{9M^2}} \sqrt{ \frac{\Lambda}{3} }\nonumber\\
&&\\
& \stackrel{\mbox{ for }Q\neq 0}{=}& \pm \frac{\left(  \frac{-K^2 x_0^2}{6MQ^2} + \frac{Kx_0}{M} - 1\right) }{ \left(  \frac{K^2 x_0^3 }{12 MQ^2} - \frac{5Kx_0}{6M} + 1\right) ^2 } \sqrt{1-\frac{8KQ^2}{9M^2} } \sqrt{ \frac{\Lambda}{3}},\nonumber
\end{eqnarray}   
where in the latter equation we use again that $K^2 = 4\Lambda Q^2$ for $g_2 = 0$.  In case $Q=0$, the $2^{nd}$ equation here reduces to $H(t_0)=\mp \sqrt{ \frac{\Lambda}{3}}$ since we can then choose $x_0=0$ as in \cite{4}, where we take the speed of light $=1$.  In fact by (4.10) since also $K=0$, $\Delta = - \frac{3\Lambda^2 M^4}{16}<0$ for $\Lambda\neq 0$ (i.e. $\Delta \neq 0$), which shows that $x_0=0$ is a non-repeated root.  In general we see that for $\Lambda >0$, $H(t_0)$ is a multiple of the Hubble constant $\sqrt{\Lambda/3}$ for the \emph{de Sitter model} with inflationary scale factor $Aexp\left( \sqrt{ \frac{\Lambda}{3}  }t \right)$, for some constant $A$.

Kraniotis and Whitehouse discuss quite a bit more (again for the case $Q=0$) regarding other possibilities also such as $\Lambda<0$ (which leads to periodic solutions), $\Lambda>0$ with $g_2\neq 0$, $g_3=0$ (for a non-Euclidean universe), or (generally) $\Delta\neq 0$, $g_2\neq 0$, $g_3\neq 0$ -- including some ``bouncing" models, and asymptotic inflationary models.

Equation (4.9) with $Q=0$ also appears in the work of N. Meures and M. Bruni on $\Lambda$CDM cosmology \cite{30}, for example.  The general solution (4.11) of course is not given there.  In fact these authors also take the curvature constant $K=0$, and thus they solve the equation
\begin{equation}
\dot\Phi^2 = \frac{2M}{\Phi} + \frac{\Lambda}{3}\Phi^2.
\end{equation}
In general, note that the equation 
\begin{equation}
\dot{y}(t)^2 = \frac{A}{y(t)}+By(t)^2
\end{equation}
has the non-elliptic solution
\begin{equation}
y(t) = \left( \frac{A}{B} \right)^{1/3}\left[  \sinh\left(  \frac{3}{2}\sqrt{B} t\right) \right] ^{2/3},
\end{equation}
for example.

Consider now the special case $\Lambda=0$.  We present solutions of (4.9) not found in Shikin \cite{29}.  In this case $f(x)=-K(r)x^2+2M(r)x-Q^2$ is quadratic.  Hence the roots of $f(x)=0$ are given by $x_0 = \left( M\pm \sqrt{M^2 - KQ^2} \right)/K$, where for convenience we write $M, K$ for $M(r), K(r)$; we also assume that $K\neq 0$.  There are $2$ possibilities to think about:  (i) $M^2\neq KQ^2$, (ii) $M^2 = KQ^2$.  If $M^2\neq KQ^2$, the roots are distinct (non-repeated).  Thus for $x_0 = \left(  M\pm \sqrt{M^2 - KQ^2}\right)/K$ the equations $-Kx_0 + M = \mp \sqrt{ M^2 - KQ^2}$, $\Lambda=0$, with equation (4.11) provide for the solutions 
\begin{eqnarray}
\Phi &=&  \frac{M\pm\sqrt{M^2 - KQ^2}}{K}\mp \frac{\sqrt{M^2 - KQ^2}}{2\left[  \wp(w+c)+K/12\right]}\nonumber\\
&&\\
t &=& \frac{M\pm\sqrt{M^2 - KQ^2}}{K} w \mp \frac{\sqrt{M^2 - KQ^2}}{2  \wp'(w_0)}\left[ \log \frac{\sigma(w+c-w_0)}{\sigma(w+c+w_0)}\right. \nonumber\\
&& \left.  + 2(w+c)\zeta(w_0)\right] + \delta\nonumber
\end{eqnarray}
for $\wp(w_0) = -K/12\neq e_1, e_2, e_3$.

To obtain quite different solutions we therefore focus on the $2^{nd}$ possibility: $M^2=KQ^2$, in which case $x_0=M/K$ is a repeated root:  $f(x) = -K(x-M/K)^2$, and the general formulas (2.2), (2.4) that lead to (4.11) do not apply.  However, equation (4.9) now assumes the form
\begin{equation}
\dot\Phi^2 = -K(\Phi - M/K)^2/\Phi^2,
\end{equation}
which has (non-parametric) solutions in terms of Lambert's $W$-function mentioned in the introduction:
\begin{equation}
\Phi(t,r) = \frac{Q^2}{M(r)} \left[  W\left(  Be^{\pm M(r)\sqrt{-K(r)}t/Q^2} \right) + 1 \right]
\end{equation}
where $B\neq 0$ is an integration constant that depends on $r$.  Note that since $M(r)^2 = K(r)Q^2$ and $K(r)\neq 0$ we also have $M(r)\neq 0$ for $Q\neq 0$, and $Q^2/M(r) = M(r)/K(r)$.  One can derive (4.21) from the general fact that the differential equation
\begin{equation}
\dot{x}(t) = b+\frac{a}{x(t)}
\end{equation}
with $a,b\neq 0$ has the solution
\begin{equation}
x(t) = \frac{a}{-b}\left[  W(Be^{-tb^2/a}) +1\right]
\end{equation}
for $B\neq0$ since $W(z)$ satisfies $W'(z) = W(z)[ z(1+W(z))]^{-1}$.

Apart from the application here, the Lambert $W$-function has been applied in a variety of disciplines ranging from statistical mechanics and quantum chemistry to enzyme kinetics, the engineering of thin films, and the physiology of vision \cite{31}.\\

\noindent {\bf Example 4.}    A dynamic correspondence between FLRW and/or Bianchi I cosmology and Bose-Einstein condensates (BECs) governed by a time-dependent, harmonic trapping potential was set up in \cite{17}.  A cosmological constant $\Lambda_d$ was present, where the spacetime dimension $d\geq 3$ was arbitrary.  Thus an extension of work of James Lidsey \cite{16} was realized.  The correspondence is presented in Tables I and II below where cosmological parameters (scale factors, scalar pressure and energy density $p_\phi, \rho_\phi$, Hubble parameters) are matched with wavepacket parameters expressed in terms of the harmonic trapping frequency $\omega(t)$ and \emph{moments} $I_j(t)$, $j=2,3,4$ (with $I_2(t)>0$) of a wave function of the \emph{Gross-Pitaevskii} (G-P) equation.  Here $t$ is ``laboratory" time that one obtains from the ``cosmic" time in the Einstein field equations.  We also assume constancy of the atomic interaction parameter in the G-P equation.  In Table I, $a(t)$ is the scale factor for the FLRW cosmological model and $H(t)\stackrel{def.}{=}\dot{a}(t)/a(t)$ is the corresponding Hubble parameter.  In Table II, $R(t)=(X_1(t)X_2(t)\cdots X_{d-1}(t))^{1/(d-1)}$ is the average scale factor with $X_j(t)$ the scale factor in the $j^{th}$ spatial direction and $H_R(t)\stackrel{def.}{=}\dot{R}(t)/R(t)$.  Also we have  $K_d\stackrel{def.}{=}8\pi G_d$  for $G_d$ the gravitational constant.

\begin{table}[h]
\centering
\parbox{11cm}{
\caption{\label{tb: BEC-FLRW}BEC $\leftrightarrow$ FLRW correspondence 
}
}
\begin{tabular}{ccccc}\hline\hline\\
$I_2$ & & $\leftrightarrow$ && $a^2$\\
$I_3$ &&$\leftrightarrow$ && $2(aH)$\\
${I_3^2}/{4I_2}$ &&$ \leftrightarrow$ &&$ H^2$\\
$\left[ (d-1) (d-2) I_4-\Lambda_d \right]/K_d$ &&$ \leftrightarrow$ && $\rho_\phi$\\
$\left[(d-2)\omega^2I_2-(d-1)(d-2)I_4+\Lambda_d\right]/K_d$ && $\leftrightarrow $&&$ p_\phi$\\
\\
\hline\hline\end{tabular}
\end{table}

\begin{table}[h]
\centering
\parbox{11cm}{
\caption{\label{tb: BEC-BI}BEC $\leftrightarrow$ Bianchi I correspondence 
}
}
\begin{tabular}{ccccc}\hline\hline\\
$I_2$ & &$ \leftrightarrow$ && $R^{2(d-1)}$\\
$I_3$ &&$\leftrightarrow$ &&$ 2(d-1)(R^{(d-1)}H_R)$\\
${I_3^2}/{4I_2}$ && $\leftrightarrow$ && $(d-1)^2H_R^2$\\
$\left[ \frac{(d-2)}{(d-1)} I_4-\Lambda_d \right]/K_d $&&$ \leftrightarrow $&& $\rho_\phi$\\
$\left[\frac{(d-2)}{(d-1)}\omega^2I_2-\frac{(d-2)}{(d-1)}I_4+\Lambda_d\right]/K_d$ &&$ \leftrightarrow$ && $p_\phi$\\
$\lambda $ &&$ \leftrightarrow $&&$ -2(d-1)K_dD/(d-2)$\\
\\
\hline\hline\end{tabular}
\end{table}

\newpage
The Tables are based on an equation of state $p_\phi = (\gamma-1)\rho_\phi$ with $\gamma>0$.  $\gamma=6/(d-1)$, for example, corresponds to a stiff perfect fluid.  The moments satisfy the \emph{conservation law}
\begin{equation}
2I_2(t)I_4(t)-I_3^2(t)/4 = \mbox{ a constant }\stackrel{def.}{=}\lambda.
\end{equation}
The two basic equations derived in \cite{17} in conjunction with Table I (with the help of (4.24)) were
\begin{equation}
\omega^2 = \frac{\alpha_0 \gamma (d-1)}{I_2^{[\gamma(d-1)+2]/2}}, \qquad \frac{\dot{I}_2^2}{4} = \frac{2\alpha_0}{I_2^{[\gamma(d-1)-2]/2}} + \frac{2\Lambda_d I_2}{(d-1)(d-2)}-\lambda
\end{equation}
for an integration constant $\alpha_0$.   These equations govern the time-dependent trapping frequency $\omega(t)$, and hence they also govern the external potential $V(r,t) = \omega(t)^2r^2/2$.  In particular, for a stiff perfect fluid the second equation here is written
\begin{equation}
\dot{I}_2(t)^2 = \frac{\alpha}{I_2(t)^2} + \frac{8\Lambda_d I_2(t)}{(d-1)(d-2)} - 4\lambda,
\end{equation}
for $\alpha=8\alpha_0$, which is equation (1.4) with $A=B=0$ there.  Instead of the parametric solution given by (2.2), (2.4), we have an alternate (simpler) parametric solution 
\begin{eqnarray}
I_2 &=& a\wp(w; g_2, g_3)+b\nonumber\\
&&\\
t&=& -a^2\zeta(w; g_2, g_3)+abw+\delta,\nonumber
\end{eqnarray}
as constructed in \cite{3}, where $a^3=(d-1)(d-2)/2\Lambda_d$, $b=\lambda(d-1)(d-2)/6\Lambda_d$, $g_2=2a(d-1)(d-2)\lambda^2/3\Lambda_d$, $g_3=2\lambda^3(d-1)^2(d-2)^2/27\Lambda_d^2-\alpha$.

The Biermann-Weierstrass solution (2.6) applies directly to provide for an explicit solution of the scale factor  $a(\eta)$ of one of the two the Friedmann equations, where eta is ``conformal" time - the other equation being a local conservation of energy equation. This is discussed in \cite{32, 33}, for example.  There energy and matter in the universe are assumed to be a perfect fluid consisting of radiation, non-relativistic matter, and a cosmological constant.  The modular property (3.1) and the  ``chain of connections" of \cite{14} discussed in section 3, and the solution (4.11) in the special Friedmann case (with $Q=0$, $x_0=0$) are also discussed in \cite{37}, for example, where a holographic description of the early universe is considered.

\vspace{1in}
\appendix{}

{\bf\Large\hspace{-.29in} Appendices}

\section{Definition of the Weierstrass phi, sigma and zeta functions}

Given the central importance of the Weierstrass phi function $\wp(w)$ for the present work we recall briefly, for the reader's convenience, its construction/ definition.  A more detailed account is available in \cite{18, 21, 34}.  

Let $\omega_1, \omega_2$ be non-zero complex numbers.  Since the imaginary parts of a non-zero complex number $z$ and its reciprocal are related by $Im \ z^{-1}=-\left( Im \ z\right) \left| z \right|^{-2}$, one has that $Im \ \omega_2/\omega_1\neq 0$ if and only if $Im \ \omega_1/\omega_2\neq 0$.  In particular we assume that $Im \ \omega_2/\omega_1>0$, which is equivalent to the assumption $Im \ \omega_1/\omega_2<0$.  The corresponding \emph{lattice} $\mathscr{L}=\mathscr{L}(\omega_1, \omega_2)$ generated by $\omega_1$ and $\omega_2$ is defined to be the set of points $\omega=m\omega_1+n\omega_2$ where $m$ and $n$ vary over the set of whole numbers.  The lattice $\mathscr{L}$ gives rise to the phi function
\begin{equation}\wp(w)\stackrel{def.}{=}\frac{1}{w^2}+\displaystyle\sum_{\omega\in\mathscr{L}-\{0\}}\left[ \frac{1}{(w-\omega)^2}-\frac{1}{\omega^2}\right]\label{eq: defnwp}\end{equation}
which is also denoted by $\wp(w; \mathscr{L})$, or by $\wp(w; \omega_1, \omega_2)$.  $\wp(w)$ is a meromorphic function, which is doubly periodic with periods $\omega_1, \omega_2$.  Thus, by definition, $\wp(w)$ is an \emph{elliptic} function.  $\wp(w)$ has double poles at $w=\omega\in\mathscr{L}$, and it satisfies the differential equation
\begin{equation}\wp'(w)^2=4\wp(w)^3-g_2(\omega_1, \omega_2)\wp(w)-g_3(\omega_1, \omega_2)\end{equation}
for \emph{invariants}
\begin{equation}
g_2(\omega_1,\omega_2)\stackrel{def.}{=}60\displaystyle\sum_{\omega\in\mathscr{L}-\{0\}} \frac{1}{\omega^4}, \ \ g_3(\omega_1, \omega_2)\stackrel{def.}{=}140 \displaystyle\sum_{\omega\in\mathscr{L}-\{0\}}\frac{1}{\omega^6}\end{equation}
where, moreover,
\begin{equation}\Delta(\omega_1, \omega_2)\stackrel{def.}{=}g_2(\omega_1, \omega_2)^3-27 g_3(\omega_1, \omega_2)^2\neq 0.\end{equation}

Conversely, it is an amazing fact that if two complex numbers $g_2$ and $g_3$ are given that satisfy the condition $g_2^3-27g_3^2\neq 0$, then there exists a pair of non-zero complex numbers $\omega_1, \omega_2$ with $Im \ \omega_2/\omega_1>0$ such that $g_2(\omega_1, \omega_2)=g_2$ and $g_3(\omega_1, \omega_2)=g_3$, for $ g_2(\omega_1, \omega_2)$ and $ g_3(\omega_1, \omega_2)$ defined in (A.3) with respect to the lattice $\mathscr{L}=\mathscr{L}(\omega_1, \omega_2)$ generated by $\omega_1$ and $\omega_2$.  Thus from $g_2$ and $g_3$ one can also construct the corresponding phi function $\wp(w; \omega_1, \omega_2)$ (according to definition (A.1)), which in this case we also denote by $\wp(w; g_2, g_3)$.

Associated with $\wp(w)$ are the Weierstrass sigma and zeta functions $\sigma(w)$ and $\zeta(w)$, respectively:
\begin{eqnarray}
\zeta'(w)\stackrel{def.}{=}-\wp(w), && \displaystyle\lim_{w\rightarrow 0} \left( \zeta(w)-\frac{1}{w}\right)\stackrel{def.}{=}0,\notag\\
&&\label{eq: defnzetasigma}\\
\notag\frac{\sigma'(w)}{\sigma(w)}\stackrel{def.}{=}\zeta(w), && \displaystyle\lim_{w\rightarrow 0} \frac{\sigma(w)}{w}\stackrel{def.}{=}1.\end{eqnarray}

\section{Theta function notation}
The notation for Jacobi theta functions varies widely and wildly from text to text, as we indicated in section 2.  There, in (2.8), we used the notation (and definitions) presented in chapter 5 of K. Chandrasekharan \cite{18}.  On the other hand, many researchers (including Kraniotis and Whitehouse \cite{4}) employ the notation in the Abramowitz-Stegun Handbook \cite{34}.  Thus for the reader's convenience we set up the following comparison:
\begin{equation}
\begin{array}{cc}
\mbox{K. Chandrasekharan}&\mbox{Abramowitz-Stegun}\\
\theta(z,\tau)&\theta_1(\pi z, q)\\
\theta_1(z,\tau)&\theta_2(\pi z, q)\\
\theta_2(z,\tau)&\theta_4(\pi z, q)\\
\theta_3(z,\tau)&\theta_3(\pi z, q)
\end{array}\nonumber
\end{equation}
for $q=e^{\pi i\tau}$, $\tau\in\Pi^+$.  Also the notation in Whittaker-Watson \cite{21} is the same as that in Abramowitz-Stegun, where the $\theta_4(z,q)$ in \cite{21} is initially denoted by $\theta(z,q)$.

The periods $\omega_1, \omega_2$ of the phi-function in Appendix A are denoted by $2\omega, 2\omega'$ in \cite{34}:  $\omega_1=2\omega$, $\omega_2=2\omega'$.  Applications of theta functions to FLRW cosmology also appear in \cite{36}, where the notation of \cite{21} is employed.

\section{On the zeros of $\wp(w; \tau)$}

The physical significance of the zeros of $\wp(w; \tau)$ has been pointed to in Example 3.  For the sake of completeness we provide a description of these zeros.

With definitions (A.3), (A.4) in mind, we introduce the normalized \emph{Eisenstein} series
\begin{equation}
E_4(\tau)\stackrel{def.}{=} \frac{3}{4\pi^4}g_2(1,\tau), \quad E_6(\tau)\stackrel{def.}{=}\frac{27}{8\pi^6}g_3(1,\tau),
\end{equation}
and normalized \emph{discriminant}
\begin{equation}
\Delta_n(\tau)\stackrel{def.}{=} \frac{\Delta(1, \tau)}{64^2 \pi^{12}} = \frac{27}{64\times 1728} \frac{\Delta(1,\tau)}{\pi^{12}}
\end{equation}
for $\tau\in\Pi^+$.  Then the zeros $z_0$ of $\wp(w; \tau)$ are given by the following explicit integral formula of M. Eichler and D. Zagier \cite{12, 24}:
\begin{equation}
z_0 = m+\frac{1}{2}+n\tau \pm\left[  \frac{\log(5+2\sqrt{6})}{2\pi i}+ 144\sqrt{6}\pi i \displaystyle\int_{\tau}^{i\infty} \frac{ (\sigma-\tau)\Delta_n(\sigma)}{E_6(\sigma)^{3/2}}d\sigma\right]
\end{equation}
for $m,n\in\mathds{Z}$, where the integral is over the vertical line $\sigma=x_0+(y_0+t)i$, $t\geq 0$, in $\Pi^+$ commencing at $\tau=x_0+y_0i$, $y_0>0$.

An alternate formula for the zeros was found by W. Duke and O. Imamoglu \cite{13}, who ``deuniformized" the Eichler-Zagier formula and expressed the zeros in terms of the classical \emph{modular invariant}
\begin{equation}
j(\tau) \stackrel{def.}{=} E_4(\tau)^3 / \Delta_n(\tau).
\end{equation}
The \emph{generalized hypergeometric series} 
\begin{equation}
{}_pF_q(a_1, \dots, a_p; b_1, \dots, b_q ;z)\stackrel{def.}{=} \displaystyle\sum_{n=0}^\infty \frac{(a_1)_n\cdots(a_p)_n}{(b_1)_n\cdots(b_q)_n} \frac{z^n}{n!}
\end{equation}
are needed, where $|z|<1$ and each $(b_k)_n\neq 0$ for the \emph{Pochhammer symbol}
\begin{eqnarray}
(a)_n& \stackrel{def.}{=} &a(a+1)(a+2)\cdots(a+n-1)=\frac{\Gamma(a+n)}{\Gamma(a)}, n\geq 1,\nonumber\\
&&\\
\nonumber (a)_0&\stackrel{def.}{=}& 1.
\end{eqnarray}
For $p=2, q=1, {}_2F_1(a_1, a_2; b_1; z)$ is the standard Gauss hypergeometric function $F(a_1, a_2; b_1; z)$.  Let
\begin{equation}
c_2 \stackrel{def.}{=} \frac{-i\sqrt{6}}{3\pi}, \quad u=u(\tau)=1- \frac{1728}{j(\tau)}=1-\frac{1728\Delta_n(\tau)}{E_4(\tau)^3},
\end{equation}
and choose the principal branch of $u^{1/4}$.  Then the zeros of $\wp(w; \tau)$ are $\pm z_0$ where $z_0$ is given by the Duke-Imamoglu formula
\begin{equation}
z_0 = \frac{1+\tau}{2} + c_2u^{1/4} \frac{{}_3F_2 (\frac{1}{3}, \frac{2}{3}, 1 ; \frac{3}{4}, \frac{5}{4}; u)}{{}_2F_1 (\frac{1}{12}, \frac{5}{12}, 1 ;  1-u)}.
\end{equation}
$u$ is given in terms of $\tau$ in (C.7).  $\tau$, conversely, is given in terms of $u$ as a quotient of hypergeometric functions:
\begin{equation}
\tau = -i + \frac{2i\sqrt{\pi}}{\Gamma\left( \frac{7}{12}\right)\Gamma\left( \frac{11}{12}\right)} \frac{{}_2F_1\left( \frac{1}{12}, \frac{5}{12}; \frac{1}{2}; u\right)} {{}_2F_1\left( \frac{1}{12}, \frac{5}{12}; 1; 1-u\right)}.
\end{equation}

In a completely different context, formula (C.8) also has an application (interestingly enough) in the work of Conte, Grundland, and Huard \cite{35}, for example, on isentropic ideal compressible fluid flow.  More specifically, (C.8) is used to conclude the boundedness (and hence the physical relevance) of a certain elliptic solution of theirs.  

Note that by (C.1), (C.2), (C.4), one also has the following expression for $j$:
\begin{equation}
j(\tau)=1728\frac{g_2(1,\tau)^3}{\Delta(1, \tau)} \stackrel{def.}{=} \frac{1728 g_2(1,\tau)^3}{g_2(1,\tau)^3 - 27 g_3(1,\tau)^2},
\end{equation}
which is the form expressed in \cite{4}, where a flat universe with cosmological constant $\Lambda\neq 0$ and $j(\tau)=0$ is considered (as we have discussed in Example 3), and where a non-Euclidean universe is also considered with $j(\tau)=1728$.  A full discussion of the modular invariant $j(\tau)$ and of Eisenstein series and modular forms, including explicit Fourier expansion formulas, is presented in \cite{26}.  In particular, we discuss there the key role that the Fourier expansion of $j(\tau)$ plays in the duality between $3$-dimensional pure gravity with a negative cosmological constant and \emph{extremal} conformal field theories with central charge $24k$, for a positive integer $k$.



\end{document}